\documentclass[sigconf]{acmart}

\usepackage{tabularx}
\usepackage{graphicx}
\usepackage{subcaption}
\usepackage{adjustbox}

\usepackage{textcomp}
\usepackage{xcolor}
\usepackage{xspace}
\usepackage{pifont}

\usepackage{algorithm}
\usepackage{algpseudocode}

\begin{document}

\def\datasetname{UAV-CAS\xspace}
\title{\datasetname: A Calibrated Digital-Twin Dataset for Intrusion Detection in UAV Swarm Networks}

\author{Sripath Mishra}
\affiliation{\institution{Purdue University} \country{USA}}
\email{mishra60@purdue.edu}

\author{Bharat Bhargava}
\affiliation{\institution{Purdue University} \country{USA}}
\email{bbshail@purdue.edu}

\author{Zizheng Liu}
\affiliation{\institution{Purdue University} \country{USA}}
\email{liu3099@purdue.edu}

\author{Shafkat Islam}
\affiliation{\institution{Purdue University Northwest} \country{USA}}
\email{islam59@pnw.edu}

\begin{abstract}
Intrusion detection systems (IDS) trained on wired-network benchmarks degrade sharply in real-world unmanned aerial vehicle (UAV) swarms, where mobility, fluctuating link quality, and decentralized routing reshape traffic distributions. Existing UAV-specific datasets also do not systematically vary these conditions, leaving no way to train or test an IDS against the very shift that defeats it. We present \datasetname, a large-scale labeled flow dataset for UAV-network intrusion detection, generated by a Containernet digital twin that is systematically calibrated against AERPAW testbed measurements. We have a four-layer calibration pipeline spanning altitude-dependent path loss, mission-specific mobility, the link-level performance chain, and end-to-end trace fidelity. \datasetname comprises 99{,}492 flows drawn from 1{,}024 configurations that span five attack families (DoS, DDoS, blackhole, wormhole, replay) and nine collaborative attack compositions. A diversity analysis shows that high-rate attacks separate from benign traffic up to an order of magnitude more strongly than in any prior benchmark, while stealth attacks deliberately blend with benign traffic. Across ten baseline IDS, binary attack detection saturates above $0.98$, confirming the dataset is learnable, whereas full attack-class identification remains hard---per-class $F_1$ ranges from near zero to $0.82$ and falls into the single digits for stealth attacks. We release the dataset, simulator, and calibration data to support reproducible UAV intrusion-detection research.
\end{abstract}


\setcopyright{none}
\acmConference{}{}{}
\settopmatter{printacmref=false}
\renewcommand\footnotetextcopyrightpermission[1]{}
\pagestyle{plain}
\maketitle


\section{Introduction}
\label{sec:intro}

Unmanned aerial vehicle (UAV) swarms are increasingly deployed for emergency response~\cite{Wenbo2020Emergency}, precision agriculture~\cite{Zeng2024FGAIDS}, environmental monitoring~\cite{Hadi2024}, and logistics operations~\cite{MURRAY201586}. These missions demand reliable multi-hop wireless communication among UAVs and ground base stations (BSs), making the network a high-value attack surface. Intrusion detection systems (IDS) trained on static, wired-network datasets---CICIDS2017~\cite{CICIDS2017}, UNSW-NB15~\cite{Moustafa20215UNSW}--- fail when deployed in UAV environments, where mobility-induced topology changes, fluctuating link quality, and decentralized routing create traffic distributions that differ from those seen during training.

The root cause is a dataset gap. Existing UAV-specific datasets---UAV Attack Dataset~\cite{Hassler2024cyberphysical}, UAV-NIDD~\cite{Hadi2025UAVNIDD}, UAVIDS-2025~\cite{UAVIDS2025}---suffer from limited scale, fixed network conditions, absent mobility modeling, or missing attack families. Current datasets do not capture \textit{collaborative attacks} where multiple adversaries coordinate different attacks (e.g., a Blackhole reroute combined with a DoS flood~\cite{chang-distributed-blackhole}), despite evidence that such compositions produce synergistic effects invisible to single-flow detectors. Furthermore, no existing UAV dataset provides a calibrated wireless channel model validated against real-world measurements. Collecting such a dataset on physical UAV hardware is prohibitively expensive and time-consuming, requiring sustained flight hours, dedicated spectrum and airspace, and the ability to safely execute attacks that physical testbeds cannot.

This paper presents \datasetname, a large-scale, labeled network traffic dataset generated from a Containernet-based UAV swarm digital twin that is systematically calibrated against measurements from the NSF AERPAW testbed~\cite{AERPAW}. We make four contributions:

\begin{enumerate}
\item \textbf{AERPAW-Calibrated Digital Twin Platform.} We design a Containernet-based UAV network emulator with a four-layer calibration pipeline---altitude-dependent path loss, mission-specific mobility dynamics, link-level performance chain, and end-to-end trace fidelity---each validated against independent AERPAW measurement campaigns (Maeng et al.~\cite{Maeng2023}, G\"urses--Sichitiu~\cite{Gurses2024}, AFAR~\cite{AFAR2025}, and AADM~\cite{AADM}). The emulator executes real Linux TCP/IP stacks inside Docker containers, producing genuine protocol artifacts.

\item \textbf{Systematic Calibration Pipeline.} We fit altitude-dependent log-distance and 3GPP TR~36.777 path loss models against AERPAW RSRP data, calibrate Gaussian--Markov mobility parameters against AFAR autonomous flight traces, validate the end-to-end link quality chain (path loss $\rightarrow$ signal-to-noise ratio (SNR) $\rightarrow$ bit error rate (BER) $\rightarrow$ packet error rate (PER) $\rightarrow$ effective throughput) against measured received signal strength (RSS) and throughput, and perform a three-way fidelity comparison showing that our digital twin achieves lower divergence from real-world traces than AERPAW's own digital twin environment.

\item \textbf{\datasetname\ Dataset.} We generate a labeled dataset spanning 1{,}024 configurations across 5 attack families (DoS, DDoS, Blackhole, Wormhole, Replay), 4 swarm sizes, 4 mission types, 2 path loss models, 2 transmit power levels, and 14 attack compositions including 9 complementary collaborative scenarios. We benchmark 10 IDS architectures on \datasetname\ and demonstrate that conventional detectors fail on collaborative attack compositions that produce flow patterns absent from any single-attack training data.

\item \textbf{Diversity Analysis.} We conduct a statistical diversity analysis using Hellinger Distance and Jensen--Shannon Divergence on inter-arrival time distributions, demonstrating that \datasetname\ exhibits significantly higher variability than all existing IDS benchmarks. This increased diversity is due to operational complexity: the same attack executed under different mobility patterns, link conditions, and topology configurations produces different flow signatures.
\end{enumerate}

All code, calibration data, configuration files, and the complete \datasetname\ dataset are publicly released to enable reproducible research.\footnote{Repository URL: https://github.com/Sripathm2/Collaborative-UAV-Dataset, Dataset Link: https://dx.doi.org/10.21227/zgrg-z865}

The remainder of this paper is organized as follows. Section~\ref{sec:related} reviews existing IDS datasets, UAV simulation platforms, wireless channel models, and dataset quality metrics. Section~\ref{sec:architecture} details the digital twin architecture, including topology, wireless channel model, mobility, attack implementation, and labeling methodology. Section~\ref{sec:calibration} describes the four-layer AERPAW calibration pipeline. Section~\ref{sec:dataset} presents the \datasetname\ dataset statistics and format. Section~\ref{sec:evaluation} evaluates calibration fidelity, dataset diversity, attack separability, baseline IDS performance, and collaborative attack detection. Section~\ref{sec:limitations} discusses scope limitations and future work.

\section{Related Work}
\label{sec:related}

\subsection{General-Purpose IDS Datasets}

Intrusion detection research has long relied on benchmark datasets collected in controlled environments. CICIDS2017~\cite{CICIDS2017} and CICIDS2018 generate realistic attack scenarios---DDoS, brute force, botnet, web exploits---over wired infrastructure, providing over eighty flow-level features. UNSW-NB15~\cite{Moustafa20215UNSW} extends attack coverage to include fuzzing, backdoors, worms, and reconnaissance using a hybrid testbed with real and synthetic traffic. CICIOT2023~\cite{Neto2023CICIOT2023} introduces wireless IoT devices and heterogeneous attack scenarios, providing broader protocol coverage but still lacking UAV-specific dynamics. Other widely used datasets---InSDN~\cite{9187858}, WIDE~\cite{WIDE-dataset}---are similarly wired-network-based. IoT-oriented datasets such as WSN-DS~\cite{ImanWSN-DS}, 5G-NIDS~\cite{Samarakoon5g}, and Bot-IoT~\cite{KORONIOTIS2019779} introduce wireless and resource-constrained elements but lack UAV-specific mobility, topology dynamics, and attack families.

All of these datasets share a critical limitation: traffic is collected on static topologies with fixed link capacity and delay, making them fundamentally unable to represent the distributional shift that UAV mobility introduces.

\subsection{UAV-Specific Datasets}

A small number of UAV-specific datasets exist. UAV Attack Dataset~\cite{Hassler2024cyberphysical} collects benign and cyber-physical attack traffic but is limited in scale and lacks UAV-to-UAV interactions. UAV Physical Dataset~\cite{UAV-Attack-Dataset} provides flight logs for GPS spoofing detection but covers only a single attack family. UAV-NIDD~\cite{Hadi2025UAVNIDD} uses three drones with varied attacks but is not parametrized over bandwidth, latency, or packet loss---conditions that change with UAV motion. UAVIDS-2025~\cite{UAVIDS2025} does not model mobility or topology changes and lacks raw packet captures needed for flow-windowed analysis. The flying ad-hoc grey hole dataset~\cite{GreyHole2025} covers a single attack family without network parameterization.

None of these datasets represent collaborative attacks, provide calibrated wireless channel models, or offer the configurability needed to evaluate IDS robustness across diverse network conditions. Table~\ref{tab:dataset_comparison} provides a structured comparison.

\subsection{Network Simulation for Dataset Generation}

Simulation platforms for UAV network research span a fidelity--feasibility spectrum. Pure packet-level simulators (NS-3~\cite{ns3}, MATLAB) model TCP/IP through mathematical abstractions but do not execute real protocol implementations---TCP retransmission behavior under load, routing convergence timing, and congestion window dynamics are approximated rather than observed. GNS3~\cite{GNS3} emulates real router firmware but lacks wireless channel modeling and mobility support. Mininet~\cite{Mininet2010} provides lightweight network emulation with real TCP/IP stacks but has no built-in mobility model, wireless propagation, or Docker container resource control support for resource-constrained node emulation.

Containernet~\cite{peuster2016containernet} extends Mininet with Docker container integration, enabling per-node resource constraints (CPU, memory) that model embedded UAV hardware. However, Containernet provides no wireless channel model; bridging this gap requires an physical-layer pipeline---which our calibration approach provides.

Physical testbeds run real UAVs with real radios but cannot safely execute adversarial experiments, are limited in swarm size by cost and airspace regulation, and cannot cover the combinatorial parameter space required for ML-based IDS evaluation. AERPAW~\cite{AERPAW}, the NSF-funded aerial experimentation platform at NC State, provides both a physical testbed and a digital twin (DT) environment. However, AERPAW's own documentation states that its DT emulation ``is not physically faithful to any particular radio or aerial environment'' and that ``no meaningful data can be usefully extracted about any physical aspect''~\cite{AERPAW_manual}. Our Containernet-based digital twin occupies a deliberate middle ground: real TCP/IP stacks inside Docker containers, with wireless channel effects applied through a calibrated physical-layer model validated against AERPAW measurements.

\subsection{Wireless Channel Modeling for UAV Communications}

UAV air-to-ground channels differ fundamentally from wired links due to altitude-dependent propagation, 3D antenna radiation patterns, and elevation-angle-dependent line-of-sight probability. The ITU-R P.1411 model~\cite{ITU_P1411} provides urban propagation predictions but does not account for UAV-specific altitude effects. The 3GPP TR~36.777 specification~\cite{3GPP_TR36777} introduces altitude-dependent path loss and shadow fading for UAV-to-ground channels in Urban Macro cells, incorporating line-of-sight probability as a function of elevation angle. The log-distance path loss model with altitude-dependent parameters has been validated against UAV measurement networks~\cite{Maeng2023,Gurses2024} and captures the dominant propagation effects without requiring ray tracing or full electromagnetic simulation.

More detailed approaches include the Enhanced Two-Ray model of Masrur and G\"uven\c{c}~\cite{Masrur2024}, which demonstrated that careful propagation modeling can close the simulation-to-real gap (18.2\,m average localization error on the AERPAW physical testbed), and ray-tracing methods that model individual multipath components at the cost of computational tractability for large-scale swarm simulation.

Our digital twin adopts the log-distance and 3GPP models as complementary options: the log-distance model is fitted directly to AERPAW RSRP measurements and captures site-specific propagation, while the 3GPP model provides a standardized reference for environments beyond the calibration site. Both models operate implicitly at the AERPAW measurement frequency ($\sim$3.5\,GHz).

\subsection{Dataset Diversity and Quality Metrics}

Evaluating dataset quality requires metrics that capture distributional differences between traffic classes and across network conditions. Hellinger Distance~\cite{Hellinger1909} and Jensen--Shannon Divergence (JSD)~\cite{Lin1991JSD} are symmetric, bounded divergence measures well-suited for comparing probability distributions of flow-level features~\cite{netdiffusion2023}. Both metrics quantify how distinguishable two distributions are: higher values indicate greater separation, implying that a dataset captures more diverse traffic patterns that an IDS must learn to discriminate.

In the context of IDS evaluation, diversity serves as a proxy for the distributional shift challenge. A dataset with low diversity (all attack flows look similar regardless of network conditions) will yield high detection accuracy that does not transfer to operational environments. Conversely, a dataset where the same attack produces measurably different flow signatures under different conditions more faithfully represents the deployment challenge. We use Hellinger Distance and JSD computed on flow packet inter-arrival time distributions as our primary diversity metrics, following~\cite{netdiffusion2023}, because packet inter-arrival times are the flow features most directly affected by changing wireless conditions.

\begin{table*}[t]
\centering
\footnotesize
\begin{tabular}{lcccccccccc}
\hline
\textbf{Dataset} & \textbf{Year} & \textbf{Scale} & \textbf{\#Attacks} & \textbf{Mobility} & \textbf{Wireless} & \textbf{Collaborative attacks} & \textbf{Labeled} & \textbf{Calibration} & \textbf{Config.} & \textbf{Open} \\
\hline
CICIDS2017~\cite{CICIDS2017}       & 2017 & 2.8M flows & 7  & \ding{55} & \ding{55} & \ding{55} & \ding{51} & N/A    & \ding{55} & \ding{51} \\
UNSW-NB15~\cite{Moustafa20215UNSW} & 2015 & 2.5M flows & 9  & \ding{55} & \ding{55} & \ding{55} & \ding{51} & N/A    & \ding{55} & \ding{51} \\
CICIOT2023 ~\cite{Neto2023CICIOT2023} & 2023 & 46M flows  & 33 & \ding{55} & Partial  & \ding{55} & \ding{51} & N/A    & \ding{55} & \ding{51} \\
UAV Attack~\cite{Hassler2024cyberphysical} & 2020 & Small & 3 & \ding{55} & \ding{51} & \ding{55} & \ding{51} & None & \ding{55} & \ding{51} \\
UAV-NIDD~\cite{Hadi2025UAVNIDD}    & 2025 & 95k flows  & 6  & \ding{55} & \ding{51} & \ding{55} & \ding{51} & None   & \ding{55} & \ding{51} \\
UAVIDS-2025~\cite{UAVIDS2025}      & 2025 & --         & 4  & \ding{55} & \ding{55} & \ding{55} & \ding{51} & None   & \ding{55} & \ding{51} \\
Grey Hole~\cite{GreyHole2025}       & 2025 & --         & 1  & \ding{51} & \ding{55} & \ding{55} & \ding{51} & None   & \ding{55} & \ding{51} \\
\hline
\textbf{\datasetname}               & 2026 & \textbf{99k flows} & \textbf{5} & \ding{51} & \ding{51} & \ding{51} & \ding{51} & \textbf{AERPAW} & \ding{51} & \ding{51} \\
\hline
\end{tabular}
\caption{Comparison of \datasetname\ with existing IDS and UAV network datasets.}
\label{tab:dataset_comparison}
\end{table*}

\section{Digital Twin Architecture}
\label{sec:architecture}

Our digital twin generates labeled UAV network traffic by simulating swarm communication inside Containernet~\cite{peuster2016containernet}, a Docker-based extension of Mininet. Each UAV and base station is instantiated as an Ubuntu~22.04 Docker container running a full Linux TCP/IP stack, producing genuine protocol behavior---TCP retransmission, congestion control, ARP resolution. The host system runs Ubuntu~22.04 with Linux kernel 6.5.0-27 on a 56-core Intel Xeon E5-2683 v3 at 2.00\,GHz with 256\,GB RAM, deployed on CloudLab~\cite{cloudlab}. To reflect onboard constraints---UAV companion computers (that are onboard the UAV and handle tasks like data collection, networking, etc.) typically use dual-core to quad-core ARM Cortex-A class processors~\cite{kangunde2021review,s21041115}---each UAV container is limited to two CPU cores, while Basestations (BS) containers are unconstrained. The 56-core host accommodates the concurrent execution of all simulated nodes: up to 20~UAV containers (each capped at 2~cores) plus BS containers, with remaining cores handling simulation orchestration, network emulation, and packet capture.

\subsection{Network Topology and Routing}
\label{sec:topology}

We adopt a switch-per-drone topology. Each node (UAV or BS) is paired with a dedicated Open vSwitch (OVS) instance treated as part of the node; all links to a node traverse its switch. BS switches are interconnected to avoid single points of failure. Nodes are placed within a $1000 \times 1000 \times [30\text{--}110]$\,m three-dimensional grid. The altitude bounds $[30, 110]$\,m are chosen to match the range of AERPAW measurement data used for path loss calibration (Section~\ref{sec:calibration_pathloss}), eliminating extrapolation.

UAVs are assigned random $(x, y, z)$ positions with a per-mission anchor point and margin from the grid boundary. BSs are placed randomly at ground level $(x, y, 0)$. All possible switch-to-switch links are pre-created and held administratively \texttt{DOWN}; the routing algorithm toggles links \texttt{UP} or \texttt{DOWN} each simulation window. While setting the link characteristics when turned UP.

\textbf{Routing algorithm.} At each window, links are established using a nearest-neighbor chain: each UAV connects to its nearest peer (UAV or BS) by Euclidean distance. If this produces disconnected components---UAV clusters unreachable from any BS---one UAV per isolated cluster is forced to connect directly to the nearest BS (bridge UAV). This power-aware bridge selection ensures full connectivity while preserving the nearest-neighbor structure.

\subsection{Wireless Channel Model}
\label{sec:channel}

Container-based emulation does not inherently model radio frequency (RF) propagation. We address this through a calibrated physical-layer pipeline that translates 3D UAV positions into per-link network impairments applied via Linux Traffic Control (\texttt{tc}) inside each container. The pipeline operates implicitly at the AERPAW measurement frequency ($\sim$3.5\,GHz); frequency-dependent propagation effects are absorbed into the fitted reference path loss $PL_0(h)$ at each altitude (Section~\ref{sec:calibration_pathloss}).

\subsubsection{Path Loss}

We support two altitude-dependent path loss models, selectable per configuration:

\textbf{Log-distance model} (fitted to AERPAW data):
\begin{equation}
PL(d, h) = PL_0(h) + 10\,n(h)\,\log_{10}\!\left(\frac{d}{d_0}\right) + X_{\sigma(h)}
\label{eq:pathloss}
\end{equation}
where $n(h)$ is the path loss exponent at altitude $h$, $PL_0(h)$ is the reference path loss at distance $d_0 = 1$\,m, and $X_{\sigma(h)} \sim \mathcal{N}(0, \sigma^2(h))$ captures altitude-dependent shadow fading. Parameters are fitted via least-squares regression against Maeng et al.~\cite{Maeng2023} RSRP measurements at five altitudes (30, 50, 70, 90, 110\,m) after antenna gain compensation, and cross-validated against independent G\"urses--Sichitiu~\cite{Gurses2024} measurements at overlapping altitudes.

\textbf{3GPP TR~36.777 Urban Macro model}: Altitude-dependent path loss following the 3GPP specification for UAV-to-ground channels~\cite{3GPP_TR36777}, with parameters derived from the same AERPAW altitude range. This model provides a standardized reference for environments beyond the calibration site.

\subsubsection{Shadow Fading}

Spatial correlation of shadow fading is modeled using the Gudmundson process~\cite{Gudmundson1991} with decorrelation distance $d_\text{corr} = 50$\,m. Each (UAV, peer) pair maintains an independent fading state with lag-1 autocorrelation:
\begin{equation}
X_\sigma[k] = \rho \cdot X_\sigma[k-1] + \sqrt{1 - \rho^2}\,\cdot\,\mathcal{N}(0, \sigma^2(h))
\end{equation}
where $\rho = e^{-\Delta d / d_\text{corr}}$ and $\Delta d$ is the distance traveled between simulation windows.

\subsubsection{Modulation and Bit Error Rate}

\begin{table}[t]
\centering

\footnotesize
\begin{tabular}{cccc}
\hline
\textbf{MCS} & \textbf{Modulation} & \textbf{Code rate} & \textbf{Data rate (Mbps)} \\
\hline
0 & BPSK   & $1/2$ & 6  \\
1 & BPSK   & $3/4$ & 9  \\
2 & QPSK   & $1/2$ & 12 \\
3 & QPSK   & $3/4$ & 18 \\
4 & 16-QAM & $1/2$ & 24 \\
5 & 16-QAM & $3/4$ & 36 \\
6 & 64-QAM & $3/4$ & 54 \\
\hline
\end{tabular}
\caption{Supported modulation and coding scheme (MCS) levels.}
\label{tab:mcs}
\end{table}

Seven modulation and coding scheme (MCS) levels are supported, following the IEEE~802.11a/g OFDM PHY~\cite{modulation} and ranging from BPSK rate-$1/2$ (6\,Mbps) to 64-QAM rate-$3/4$ (54\,Mbps), as listed in Table~\ref{tab:mcs}. In \textit{adaptive} mode, the highest-rate MCS whose required SNR threshold is met with a 3\,dB margin is selected per link per window. BER is computed using the Q-function formulation from Proakis~\cite{Proakis2008DigitalComm}:
\begin{equation}
\text{BER}_\text{M-QAM} = \frac{4}{\log_2 M}\left(1 - \frac{1}{\sqrt{M}}\right) Q\!\left(\sqrt{\frac{3\,\text{SNR}_\text{lin}}{M-1}}\right)
\end{equation}
The packet error rate (PER) for a packet of $L$ bits is $\text{PER} = 1 - (1 - \text{BER})^L$.

\subsubsection{Link Impairment Application}
Per-link parameters (bandwidth, delay, packet loss) are derived from the path loss $\rightarrow$ SNR $\rightarrow$ MCS $\rightarrow$ BER $\rightarrow$ PER chain and are applied \emph{per link}: when a link is brought UP for a simulation window, its computed bandwidth, delay, and loss are installed on that link via \texttt{tc netem}, using the channel characteristics of that specific (node, peer) pair. Each link is therefore impaired according to its own geometry, rather than sharing a single per-node value.

The \texttt{netem} qdisc is installed through Containernet's \texttt{TCLink} support. We verified placement via \texttt{docker exec mn.<node> tc qdisc show dev <iface>}: links carry the \texttt{netem} qdisc with their per-window parameters, while interfaces that only switch traffic show \texttt{noqueue} (OVS handles switching, unimpaired).

Because impairment is attached per link, a packet that traverses multiple links accumulates each link's delay and loss once, in sequence---the physically correct behavior for a multi-hop path---with no link shaped more than once per window.

\subsubsection{Environmental Parameterization}
\label{sec:env_param}

The configuration axes of \datasetname\ are chosen to present distinct physical mechanisms that alter flow-level traffic characteristics:

\textbf{TX power} (10 vs.\ 30\,dBm) controls the SNR operating point. At 10\,dBm, links at moderate distances may fall below the 64-QAM threshold, triggering lower MCS selections that reduce effective bandwidth and increase PER. At 30\,dBm, most links operate at the maximum 54\,Mbps rate. This axis produces diversity in throughput and packet loss distributions even when topology and attacks are held constant.

\textbf{Mission type} (spiral, grid, hover-transit, random) determines how UAV positions evolve over time, directly affecting topology change rate, link duration distributions, and the frequency of bridge-UAV reassignments. Spiral and grid missions produce predictable, gradual link evolution; random walks produce abrupt topology changes; hover-transit produces alternating stable (hover) and dynamic (transit) phases.

\textbf{Swarm size} (5, 10, 15, 20 drones) controls network density. Sparse swarms (5~drones) produce star topologies where most UAVs connect directly to a BS; dense swarms (20~drones) produce multi-hop chains where bridge UAVs become critical, congestion patterns emerge, and collaborative attacks have more surface area.

\textbf{Path loss model} (log-distance vs.\ 3GPP) produces different SNR distributions for the same physical geometry, exercising the sensitivity of flow features to wireless channel assumptions. This axis directly tests whether IDS models are robust to propagation model uncertainty.

These axes interact multiplicatively: a 5-drone spiral at 30\,dBm produces a stable, high-SNR star topology, while a 20-drone random walk at 10\,dBm produces a volatile, SNR-stressed multi-hop mesh. This combinatorial interaction is what drives the statistical diversity documented in Section~\ref{sec:eval_diversity}.

\subsection{Mobility Model}
\label{sec:mobility}

UAV positions evolve according to mission-specific mobility models calibrated against AFAR autonomous flight traces~\cite{AFAR2025} (Section~\ref{sec:cal_mobility}). Four mission types are supported:

\textbf{Spiral.} An Archimedean spiral at fixed altitude, modeling circular area search. Fitted parameters: initial radius $r_0$, pitch (inter-loop spacing), speed, and altitude.

\textbf{Grid.} A boustrophedon (lawn-mower) pattern at fixed altitude, modeling systematic area coverage. Fitted parameters: bounding box $(x_0, y_0, x_1, y_1)$, lane spacing, speed, and altitude (clamped identically to spiral).

\textbf{Hover-transit.} A sequence of waypoints with hover periods at each, modeling data-mule or inspection missions. Altitude forced to 50\,m to keep UAVs within the operational envelope during hover periods.

\textbf{Random.} A Gauss--Markov random walk~\cite{gaussian-markov-model} with fitted velocity parameters including a negative vertical mean (reflecting the tendency of UAVs to descend). The walk reflects on all six faces of the bounding box.

\subsection{Attack Implementation}
\label{sec:attacks}

Five attack families are implemented, each using tools executing inside Docker containers (Table~\ref{tab:attack_taxonomy}):

\textbf{DoS.} A single attacker UAV launches SYN floods against a randomly selected victim using \texttt{hping3} with configurable inter-packet interval. Default is set at \texttt{rate=1000} ($\sim$1{,}000 packets/s), a low-rate attack intensity that blends with benign traffic more realistically than high-rate floods and better represents operational threat scenarios where attackers attempt to evade detection.

\textbf{DDoS.} Identical to DoS but with 2 or more randomly selected attacker UAVs targeting a single victim, representing a \textit{synchronized} collaborative attack.

\textbf{Blackhole.} An OVS flow-drop rule is installed on the victim's switch. The link remains \texttt{UP}---ARP functions normally, but data frames are silently dropped. This produces a realistic blackhole signature: the victim appears connected but experiences progressive throughput degradation.

\textbf{Wormhole.} A Layer~2 reroute through the attacker's switch. The victim's traffic is redirected at the switch level to traverse the attacker's switch before reaching the BS. IP headers remain untouched---no attacker$\leftrightarrow$victim IP flow exists. The observable effect is increased latency and reduced throughput for victims due to the additional hop.

\textbf{Replay.} A Scapy-based packet capture and replay attack executing inside the attacker container. Packets are captured on the attacker's \texttt{eth0} interface, buffered, and replayed with modified TTL and sequence numbers. Replayed packets retain original source and destination IPs, producing elevated traffic volume at the victim without a distinct attacker IP signature. Two replay variants are included with different operational profiles: (i)~conservative (buffer=50pkts, 200\,ms delay, 10\,pps, TTL$-$5, sequential increment) and (ii)~aggressive (buffer=100pkts, 100\,ms delay, 20\,pps, TTL$-$3, random sequence), enabling evaluation of detection sensitivity to replay intensity.

\begin{table}[t]
\centering
\footnotesize
\begin{tabular}{p{1.3cm}p{2.5cm}p{2.0cm}p{1.5cm}}
\hline
\textbf{Attack} & \textbf{Mechanism} & \textbf{Parameters} & \textbf{Signature} \\
\hline
DoS      & \texttt{hping3} SYN flood           & Rate ($\mu$s interval)                & IP-level flood \\
DDoS     & Multi-source SYN flood              & Rate, \#attackers                      & Multi-src flood \\
Blackhole & OVS flow-drop rule                 & Duration                               & Silent drop \\
Wormhole & L2 switch reroute                   & Duration                               & Extra hop \\
Replay   & Scapy capture + replay              & Buffer, delay, rate, TTL, seq          & Volume spike \\
\hline
\end{tabular}
\caption{Attack taxonomy with implementation details and configurable parameters.}
\label{tab:attack_taxonomy}
\end{table}

\subsubsection{Collaborative Attack Compositions}

\datasetname\ includes both single-attack scenarios and collaborative compositions where two attacks execute simultaneously. All collaborative compositions in \datasetname\ are \textit{complementary}: attackers execute different attack types that produce synergistic effects. For example, Blackhole+DoS combines silent packet dropping with flooding, making the blackhole harder to detect amid the DoS noise; Wormhole+DoS reroutes traffic through the attacker while simultaneously flooding the network; Blackhole+Replay combines silent dropping with traffic inflation, producing contradictory flow-level signals.

Table~\ref{tab:collab_taxonomy} enumerates all 16 attack compositions in the campaign.

\begin{table}[t]
\centering
\footnotesize
\begin{tabular}{llc}
\hline
\textbf{\#} & \textbf{Composition} & \textbf{Type} \\
\hline
1  & Benign                                   & --- \\
\hline
2  & DoS ($=1000$)                            & Single \\
3  & DDoS ($=1000$)                           & Single \\
4  & Blackhole                                & Single \\
5  & Wormhole                                 & Single \\
6  & Replay (conservative)                    & Single \\
7  & Replay (aggressive)                      & Single \\
\hline
8  & Blackhole + DoS                          & Complementary \\
9  & Blackhole + DDoS                         & Complementary \\
10 & DoS + Wormhole                           & Complementary \\
11 & DDoS + Wormhole                          & Complementary \\
12 & Blackhole + Wormhole                     & Complementary \\
13 & Blackhole + Replay (cons.)               & Complementary \\
14 & DDoS + Replay (cons.)                    & Complementary \\
15 & DoS + Replay (cons.)                     & Complementary \\
16 & Replay (cons.) + Wormhole                & Complementary \\
\hline
\end{tabular}
\caption{Attack compositions in \datasetname. All collaborative compositions are complementary.}
\label{tab:collab_taxonomy}
\end{table}

\subsection{Traffic Generation}
\label{sec:traffic}

Benign traffic is generated by running an NGINX~\cite{nginx} server (single worker) on each UAV container, which continuously transmits still images~\cite{drones6070161} to BSs via HTTP, emulating sensing and telemetry pipelines common in UAV surveillance and inspection missions. FTP transfers are distributed round-robin across BSs: drone $i$ transmits to $\text{BS}_{i \bmod |\text{BS}|}$, ensuring balanced load.

\subsection{Labeling Methodology}
\label{sec:labeling}

Ground truth labels are generated deterministically from the simulation configuration. Each run produces an \texttt{attack\_details} log recording: the attack type(s), attacker and victim IP addresses, attack start time, attack duration, and all physical-layer parameters (SNR, path loss, MCS selection) for each simulation window. The log also records \texttt{tc} parameters applied to each container and routing topology changes.

Flows are labeled by matching the 5-tuple (source IP, destination IP, source port, destination port, protocol) against the attacker/victim IPs from the attack log. A flow is labeled as an attack type if either its source or destination IP matches a known attacker or victim for that attack. Flows not matching any attack record are labeled \texttt{Benign}.

For collaborative attack compositions, a flow may carry labels for multiple simultaneous attacks (e.g., a victim targeted by both blackhole and DoS receives both labels). This multi-label structure supports both binary (attack vs.\ benign) and multi-class evaluation.

\subsection{Configuration Space}
\label{sec:config_space}

\begin{table}[t]
\centering
\footnotesize
\begin{tabular}{ll}
\hline
\textbf{Parameter} & \textbf{Values} \\
\hline
Attack composition   & 16 combinations (Table~\ref{tab:collab_taxonomy}) \\
Number of drones     & 5, 10, 15, 20 \\
Number of BSs        & 2 \\
Payload type         & Image \\
Path loss model      & Log-distance (fitted), 3GPP \\
Modulation           & Adaptive \\
Mission type         & Spiral, Grid, Hover-transit, Random \\
TX power (dBm)       & 10, 30 \\
Noise floor           & $-95$\,dBm \\
\hline
\end{tabular}
\caption{Configuration parameter space. Total: $16 \times 4 \times 1 \times 1 \times 2 \times 1 \times 4 \times 2 \times 1 = 1{,}024$ configurations.}
\label{tab:config_space}
\end{table}

The full Dataset spans the Cartesian product of the parameter axes in Table~\ref{tab:config_space}. Each configuration is encoded as a hyphen-separated string, enabling deterministic replay. The simulation executes 12 mobility windows of 5 seconds each, producing approximately 60 seconds of data collection per configuration plus overhead for topology setup, \texttt{pingAll} connectivity verification, FTP initialization, and attack dispatch. Configurations are deterministically shuffled (seeded pseudorandom permutation) before distribution across compute nodes to ensure that partial campaign completion yields proportional coverage of all parameter axes.

\subsection{Traffic Capture and Flow Extraction}
\label{sec:capture}

Network taps at every switch capture all packets into pcap files. Pcaps are processed by \texttt{tshark} to extract per-packet fields: \texttt{frame time\_epoch}, \texttt{frame len}, \texttt{ip src}, \texttt{ip dst}, \texttt{tcp src port}, \texttt{tcp dst port}, \texttt{udp src port}, \texttt{udp dst port}, and \texttt{tcp flags}. Packets are aggregated into bidirectional flows using the 5-tuple with no timeout---all packets sharing a 5-tuple within a configuration run belong to the same flow. Flows with fewer than 10 packets are dropped to filter noise, and flows are capped at 100{,}000 packets.

Because packet capture uses \texttt{tcpdump -i any}, encapsulated packets may produce comma-separated multi-layer IP addresses; our processing takes the innermost (actual endpoint) IP.

\section{Calibration Pipeline}
\label{sec:calibration}

The digital twin employs a four-layer calibration pipeline, where each layer tunes simulation parameters against measurements from the NSF AERPAW testbed~\cite{AERPAW}. Fidelity is assessed using distributional divergence (Hellinger Distance, Jensen--Shannon Divergence), temporal correlation (Pearson coefficient), error metrics (RMSE, median absolute error), and the Kolmogorov--Smirnov test. The pipeline draws on four complementary AERPAW measurement datasets:

\begin{itemize}
\item \textbf{Maeng et al.}~\cite{Maeng2023}: RSRP at 3.51\,GHz across five UAV altitudes (30--110\,m) with measured 3D antenna radiation patterns, enabling decoupling of antenna gain from propagation loss.
\item \textbf{G\"urses--Sichitiu}~\cite{Gurses2024}: Wideband channel sounding at 3.3\,GHz across three altitudes and three bandwidths at five ground nodes, enabling independent cross-validation.
\item \textbf{AFAR}~\cite{AFAR2025}: 14 real-world autonomous search flights with GPS, velocity, orientation, RSS, and RSQ, totaling $\sim$300{,}000 samples across diverse mobility patterns including spiral sweeps, grid searches, and Bayesian optimization-driven trajectories, plus 15 paired AERPAW DT datasets.
\item \textbf{AADM}~\cite{AADM}: Multi-modal data-mule missions with SNR and throughput traces across multiple BSs, with paired DT and real-world evaluations.
\end{itemize}

\subsection{Layer 1: Altitude-Dependent Path Loss Model}
\label{sec:calibration_pathloss}

Real-world UAV channels exhibit altitude-dependent behavior---ground reflection, 3D antenna radiation pattern effects, and elevation-dependent shadow fading. We fit the log-distance model (Eq.~\ref{eq:pathloss}) via least-squares regression against Maeng et al.\ RSRP data after antenna gain compensation at each of the five calibration altitudes.

\textbf{Cross-validation.} We cross-validate fitted parameters against the independent G\"urses--Sichitiu~\cite{Gurses2024} measurements at overlapping altitudes (40, 70, 100\,m) (Table~\ref{tab:pathloss_params}). This cross-validation across two campaigns with different hardware (USRP B205mini vs.\ B210), waveforms (LTE vs.\ channel sounder), and bandwidths (1.25, 2.5, 5\,MHz) provides confidence that the fitted parameters generalize beyond a single measurement artifact.

\begin{table}[t]
\centering
\footnotesize
\begin{tabular}{cccc}
\hline
\textbf{Altitude (m)} & $n(h)$ & $PL_0(h)$ (dB) & $\sigma(h)$ (dB) \\
\hline
30  & 1.541 & 15.23 & 3.70 \\
50  & 1.593 & 13.65 & 3.73 \\
70  & 1.849 & 4.01 & 4.08 \\
90  & 1.881 & 3.16 & 4.48 \\
110 & 1.860 & 3.37 & 4.46 \\
\hline
\end{tabular}
\caption{Fitted path loss parameters from Maeng et al.\ RSRP data.}
\label{tab:pathloss_params}
\end{table}

\subsection{Layer 2: Mobility and Topology Calibration}
\label{sec:cal_mobility}

We calibrate Gaussian--Markov parameters ($\alpha$, $\bar{v}$, $\sigma^2_v$) and mission-driven trajectory models against AFAR and AADM traces (Table~\ref{tab:mobility_params}). For each real-world trace, we extract velocity distributions, heading change-rate distributions, altitude profiles, spatial visit frequency maps, and link lifetime distributions computed via the Layer~1 calibrated path loss model.

For Gaussian--Markov, we fit parameters by minimizing JSD against velocity distributions. For mission-driven patterns (spiral, grid, hover-transit), we match spatial visit frequency and heading change-rate distributions.

\begin{table}[t]
\centering

\footnotesize
\begin{tabular}{lcccc}
\hline
\textbf{Mission} & $\alpha$ & $\bar{v}$ (m/s) & $\sigma^2_v$ & \textbf{Altitude (m)} \\
\hline
Spiral & -- & 0.570 & -- & 25.96 (clamped to 30) \\
Grid & -- & 9.960 & -- & 26.06 (clamped to 30) \\
Hover-transit & -- & 9.962 & -- & 37.53 (forced to 50) \\
Random & 0.629 & 2.501 & 3.713 & Varies \\
\hline
\end{tabular}
\caption{Fitted mobility parameters per mission type from AFAR/AADM traces.}
\label{tab:mobility_params}
\end{table}

\subsection{Layer 3: Link-Level Performance Validation}
\label{sec:cal_linklevel}

The digital twin computes link quality through the chain. Layer~3 validates this end-to-end chain by replaying real-world AFAR and AADM trajectories through the simulator and comparing simulated RSS, SNR, and throughput against measured values.

We evaluate two propagation variants: our Layer~1 altitude-dependent log-distance model and the 3GPP TR~36.777 Urban Macro model. We report median RSS error and the Pearson correlation of the SNR time series. Replaying 30 AADM/AFAR testbed flights (1.8M paired link samples) through the calibrated log-distance variant yields a median absolute RSS error of 13.1\,dB after correcting a $-6.0$\,dB mean cross-sensor calibration bias (18.3\,dB uncorrected); the error shrinks with altitude, from 18.5\,dB below 40\,m to 10.7\,dB in the 60--80\,m band. SNR time-series tracking is weak but statistically significant (median Pearson $r \approx 0.09$ across flights, up to $0.19$ on the best-tracked flights, $p \approx 0$): the model reproduces coarse distance-driven trends but not the fine-grained temporal fluctuations of the measured link, an expected consequence of emulating rather than physically propagating the channel. We do not report throughput RMSE because none of the replayed AERPAW campaigns (AADM, AFAR, Maeng et al.) include measured end-to-end throughput.

\subsection{Layer 4: End-to-End Trace Fidelity}
\label{sec:cal_e2e}

Using paired DT and real-world traces from AFAR and AADM, we perform a three-way fidelity comparison. We configure our Containernet digital twin to match AERPAW experimental parameters, replay real-world trajectories, and compute per-feature Hellinger Distance across three pairs:

\begin{itemize}
\item[(a)] Our DT $\leftrightarrow$ Real world
\item[(b)] AERPAW DT $\leftrightarrow$ Real world
\item[(c)] Our DT $\leftrightarrow$ AERPAW DT
\end{itemize}

Features compared include RSS, inter-arrival time, packet size, link duration, and topology change rate distributions.

\textit{The simulator reproduces its trajectory source's RSS distribution far more closely than two independent real-measurement campaigns reproduce each other.} On the winning log-distance$+$shadow variant, the Hellinger Distance between simulated and AFAR-derived RSS is 0.33, well below the 0.73 \emph{cross-sensor ceiling} between the two real campaigns (Maeng~$\leftrightarrow$~AFAR). The simulation-to-Maeng distance is higher (0.92), but this is dominated by a population-sampling difference---Maeng's fixed anchor-altitude bench measurements sample the channel very differently from full AFAR flights---rather than by a propagation-fidelity failure. This variant minimizes total simulation-to-real divergence across the four calibration sweeps.

\section{\datasetname\ Dataset}
\label{sec:dataset}

\begin{table}[t]
\centering
\footnotesize
\begin{tabular}{lc}
\hline
\textbf{Class} & \textbf{Number of Flows} \\
\hline
Benign        & 58,589 \\
DoS           & 1,223 \\
DDoS          & 7,316 \\
Blackhole     & 12,116 \\
Wormhole      & 10,862 \\
Replay        & 2,025 \\
Collaborative & 7,361 \\
\hline
\textbf{Total} & \textbf{99,492} \\
\hline
\end{tabular}
\caption{\datasetname\ dataset statistics.}
\label{tab:dataset_stats}
\end{table}

Table~\ref{tab:dataset_stats} summarizes the class distribution. Each flow record contains the following per-packet fields extracted by \texttt{tshark}: timestamp (\texttt{frame.time\_epoch}), frame length (\texttt{frame.len}), source and destination IP addresses, source and destination TCP/UDP ports, and TCP flags. From these raw fields, flows are represented as ordered sequences of (timestamp, size, flags) tuples, enabling downstream extraction of inter-arrival time distributions, packet size distributions, burst statistics, and flag-based behavioral features.

\subsection{Intended Use Cases}
\label{sec:use_cases}

\datasetname\ is designed to support multiple evaluation paradigms:

\textbf{Binary detection} (attack vs.\ benign): All five attack families are collapsed into a single ``attack'' class. This setting evaluates whether an IDS can distinguish any anomalous activity from normal UAV operations.

\textbf{Multi-class classification} (benign + 5 attack types): Each flow is assigned one of six labels. This setting evaluates whether an IDS can not only detect but also identify the attack family---critical for selecting appropriate mitigation responses.

\textbf{Collaborative attack detection}: Configurations with two simultaneous attacks produce flows with multi-label ground truth. This setting evaluates whether detectors trained on single-attack data can recognize emergent collaborative patterns.

\textbf{Cross-condition generalization}: By partitioning \datasetname\ along configuration axes (e.g., train on spiral missions, test on random; train on 5-drone swarms, test on 20-drone), researchers can evaluate IDS robustness to distributional shift---the core challenge in operational UAV deployment.

We recommend a 70/10/20 train/validation/test split with standard scaling applied to all features. For cross-condition experiments, we recommend holding out entire configuration axis values (e.g., all random-mission configs) rather than random sampling, to ensure the test set contains genuinely unseen network conditions.

\section{Evaluation}
\label{sec:evaluation}

We evaluate the \datasetname\ digital twin and dataset along five axes: calibration fidelity (Sections~\ref{sec:eval_pathloss}--\ref{sec:eval_e2e}), dataset diversity (Section~\ref{sec:eval_diversity}), attack separability (Section~\ref{sec:eval_separability}), baseline IDS benchmarking (Section~\ref{sec:eval_baselines}), and collaborative attack detection (Section~\ref{sec:eval_collab}).

\subsection{Evaluation Setup}
\label{sec:eval_setup}

\textbf{Calibration experiments} use AERPAW measurement datasets as ground truth (Section~\ref{sec:calibration}). \textbf{Input}: Maeng et al.\ RSRP data, G\"urses--Sichitiu channel sounding data, AFAR/AADM paired real-world and DT traces. \textbf{Output}: RMSE, MAE, Pearson correlation, Hellinger Distance, JSD, and KS-test statistics for each calibration layer. \textbf{Baselines}: Enhanced Two-Ray model~\cite{Masrur2024} and 3GPP TR~36.777 model (for path loss); AERPAW's own DT traces (for end-to-end fidelity).

\textbf{Diversity experiments} use the full \datasetname\ dataset alongside four benchmark datasets: CICIDS2017~\cite{CICIDS2017}, UNSW-NB15~\cite{Moustafa20215UNSW}, CICIOT2023~\cite{Neto2023CICIOT2023}, and UAV-NIDD~\cite{Hadi2025UAVNIDD}. \textbf{Input}: per-flow inter-arrival time sequences from each dataset. \textbf{Output}: Hellinger Distance and Jensen--Shannon Divergence between benign and attack distributions within each dataset. \textbf{Metric interpretation}: higher scores indicate greater within-dataset diversity, implying that the same attack manifests differently across network conditions.

\textbf{Baseline IDS experiments} evaluate 10 IDS architectures on \datasetname: \textbf{Logistic Regression} (LR)~\cite{Hassler2024cyberphysical}, \textbf{Stochastic Gradient Descent} (SGD)~\cite{Hassler2024cyberphysical}, \textbf{Random Forest} (RF)~\cite{Hassler2024cyberphysical}, \textbf{MLP}~\cite{Hassler2024cyberphysical}, \textbf{1D-CNN}~\cite{Hassler2024cyberphysical}, \textbf{LSTM}~\cite{Hassler2024cyberphysical}, \textbf{LightGBM}~\cite{Silva2023deepswarm}, \textbf{ConvNet}~\cite{Haija22}, \textbf{TinyML}~\cite{Wu2024tml}, and \textbf{CNN-BiLSTM}~\cite{Sinha2020BIRNN}. \textbf{Input features}: aggregated flow-level statistics computed from each flow's packet sequence---duration, total forward/backward packets, byte lengths (total, mean, std, min, max), inter-arrival time statistics (mean, std, min, max), and TCP flag counts (SYN, ACK, FIN, RST, PSH, URG). All inputs are standard-scaled. \textbf{Output}: 6-class label (Benign, DoS, DDoS, Blackhole, Wormhole, Replay). \textbf{Data split}: 70\% training, 10\% validation, 20\% testing, stratified by class. Neural baselines are trained for 50 epochs with Adam optimizer (learning rate $10^{-3}$, batch size 128); architectures follow original reference implementations without further hyperparameter tuning. \textbf{Metrics}: weighted $F_1$-score for overall performance and per-class $F_1$ for attack-specific analysis. The same experimental protocol is applied identically across CICIDS2017, UNSW-NB15, CICIOT2023, and \datasetname to enable fair cross-dataset comparison.

\textbf{Collaborative attack experiments} evaluate whether baselines trained on single-attack configurations can detect collaborative compositions. \textbf{Input}: the same feature set and baselines as above. \textbf{Training data}: flows from single-attack configurations only (compositions 1--7 in Table~\ref{tab:collab_taxonomy}). \textbf{Test data}: flows from collaborative configurations (compositions 8--16). \textbf{Output}: per-composition $F_1$-score revealing which collaborative patterns evade single-attack-trained detectors. This protocol directly evaluates the emergent detection challenge that motivates \datasetname's collaborative attack design.

\textbf{Hardware}: All experiments are executed on the same CloudLab host described in Section~\ref{sec:topology} (56-core Intel Xeon E5-2683 v3 at 2.00\,GHz, 256\,GB RAM, Ubuntu~22.04); IDS training is CPU-only.

\subsection{Path Loss Calibration Fidelity}
\label{sec:eval_pathloss}

\textbf{The fitted log-distance model achieves the lowest overall error and the best held-out cross-validation, beating the 3GPP reference by $\sim$2.9\,dB RMSE on Maeng test data and by $\sim$6.6\,dB on G\"urses--Sichitiu cross-validation, while remaining competitive with an Enhanced Two-Ray model fitted to the same data.}

Table~\ref{tab:pathloss_eval} reports RMSE, MAE, and Pearson correlation per altitude on Maeng et al.\ held-out test data (17{,}043 samples; 70/30 train/test split per altitude), aggregate overall metrics, and a fully held-out cross-validation against G\"urses--Sichitiu measurements (3{,}131 samples) collected with different hardware (USRP B210 vs.\ B205mini), waveform (channel sounder vs.\ LTE), and bandwidth.

\begin{table*}[t]
\centering
\footnotesize
\begin{tabular}{lccc|ccc|ccc}
\hline
 & \multicolumn{3}{c|}{\textbf{Log-Distance (fitted)}} & \multicolumn{3}{c|}{\textbf{Two-Ray (fitted)}} & \multicolumn{3}{c}{\textbf{3GPP TR~36.777}} \\
\textbf{Altitude (m)} & RMSE & MAE & $r$ & RMSE & MAE & $r$ & RMSE & MAE & $r$ \\
\hline
30   & \textbf{3.69} & \textbf{2.21} & \textbf{0.76} & 4.87          & 3.21          & 0.75          & 7.21 & 5.32 & 0.73 \\
50   & \textbf{4.00} & \textbf{2.63} & 0.70          & 4.59          & 3.03          & \textbf{0.73} & 7.33 & 5.64 & 0.73 \\
70   & 4.42          & 2.45          & 0.72          & \textbf{4.13} & \textbf{2.35} & \textbf{0.75} & 7.30 & 6.00 & 0.75 \\
90   & 4.87          & 3.11          & 0.66          & \textbf{4.72} & \textbf{2.75} & \textbf{0.70} & 6.80 & 4.87 & 0.70 \\
110  & \textbf{4.60} & 2.94          & 0.66          & 4.72          & \textbf{2.69} & 0.69          & 7.87 & 6.91 & \textbf{0.69} \\
\hline
\textbf{Overall (Maeng)}                & \textbf{4.40} & \textbf{2.69} & \textbf{0.70} & 4.60 & 2.75 & 0.67 & 7.32  & 5.79  & 0.65 \\
\textbf{Cross-val (G\"urses-Sichitiu)} & \textbf{7.04} & \textbf{5.90} & \textbf{0.39} & 7.21 & 6.49 & 0.37 & 13.63 & 13.77 & 0.38 \\
\hline
\end{tabular}
\caption{Path-loss model accuracy on Maeng et al.\ held-out test set per altitude, plus aggregated overall metrics and held-out cross-validation against independent G\"urses--Sichitiu measurements. RMSE / MAE in dB (lower is better); Pearson $r$ unitless (higher is better).}
\label{tab:pathloss_eval}
\end{table*}

\textbf{The largest gains over the 3GPP reference appear at low altitudes (30--50\,m) where ground reflection and elevation-dependent radiation pattern effects are strongest.} At 30\,m, our fitted log-distance model reduces RMSE from 7.21\,dB (3GPP) to 3.69\,dB---a 49\% reduction---and MAE from 5.32\,dB to 2.21\,dB. The Enhanced Two-Ray model is competitive at 70--90\,m where the two-ray ground-reflection geometry matches the measurement environment, but log-distance retains the overall edge thanks to its lower error at 30\,m and 50\,m, where the bulk of test samples lies.

\textbf{The $\sim$6.6\,dB cross-validation gap between log-distance and 3GPP (7.04\,dB and 13.63\,dB) confirms that the fitted parameters generalize across measurement campaigns, hardware, and waveforms} rather than memorizing artifacts of the Maeng data alone. Cross-validation Pearson correlation is uniformly low ($r \approx 0.38$) across all three models, reflecting the inherent variability of the G\"urses--Sichitiu campaign (1--5\,MHz channel sounder over a different deployment site); the discriminating signal is in the absolute error, not in correlation.

\begin{figure*}[t]
\centering
\begin{subfigure}{0.51\textwidth}
  \centering
  \includegraphics[width=\linewidth]{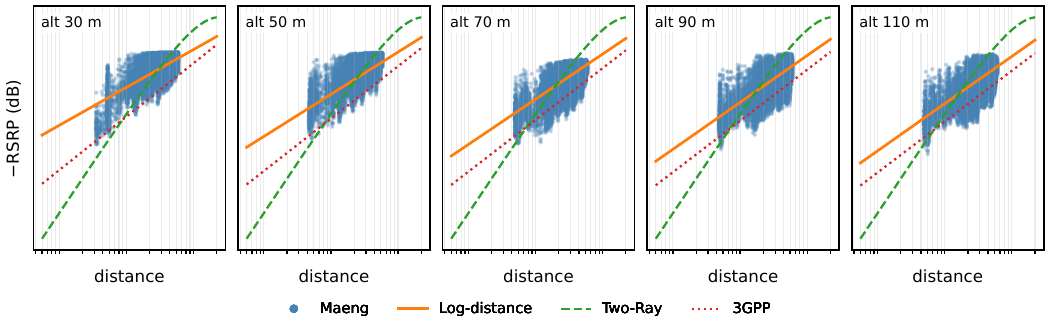}
  \caption{Path loss vs.\ distance at five AERPAW altitudes.}
  \label{fig:pathloss_curves}
\end{subfigure}
\hfill
\begin{subfigure}{0.45\textwidth}
  \centering
  \includegraphics[width=\linewidth]{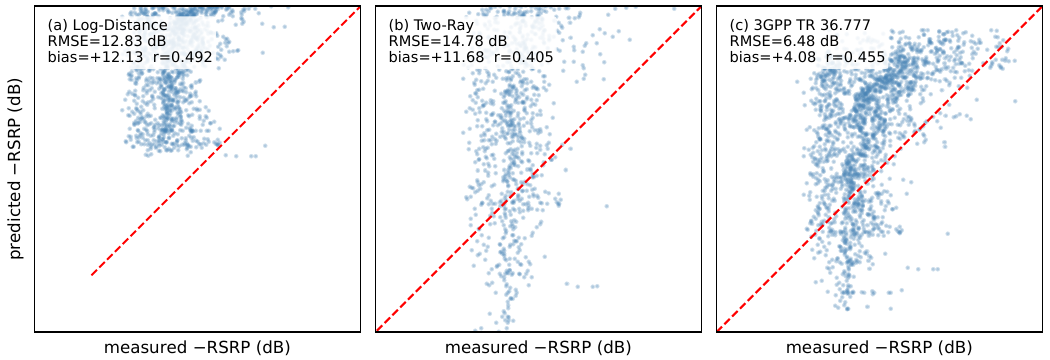}
  \caption{Cross-validation against independent G\"urses--Sichitiu measurements.}
  \label{fig:pathloss_crossval}
\end{subfigure}
\caption{Path-loss model fidelity. (\subref{fig:pathloss_curves})~Fitted log-distance (solid), fitted Two-Ray (dashed), and 3GPP TR~36.777 reference (dotted) against Maeng et al.\ measurements (points) at five altitudes; bottom row shows residuals with $\pm\sigma(h)$ bands. (\subref{fig:pathloss_crossval})~Held-out predictions vs.\ G\"urses--Sichitiu independent measurements at 40, 70, and 100\,m.}
\label{fig:pathloss}
\end{figure*}

Figure~\ref{fig:pathloss_curves} plots simulated path loss against Maeng et al.\ measurements at five altitudes with the Two-Ray and 3GPP references overlaid; the residual subplot confirms that errors are centered at zero with altitude-dependent variance matching the fitted $\sigma(h)$. Figure~\ref{fig:pathloss_crossval} shows the cross-validation against G\"urses--Sichitiu independent measurements.

\subsection{Mobility Calibration Fidelity}
\label{sec:eval_mobility}

\textbf{Fitted mission-specific mobility parameters reproduce real AFAR speed distributions to within JSD~$\leq$~0.66 on every held-out flight, with the Gauss--Markov random walk achieving the closest match (JSD~$=$~0.16).}

Table~\ref{tab:mobility_eval} reports per-mission velocity-distribution Jensen--Shannon distance (JSD on speed histograms; lower is better) and spatial visit-frequency Kullback--Leibler divergence (KL on $(x,y)$ occupancy maps; lower is better) on AFAR test flights that were not used during fitting. Each mission's parameters were fit on a hand-coded subset of AFAR flights (3--5 flights per mission, 70\% training fraction, 0.2\,s sample step) and tested on 1--2 held-out flights.

\begin{table}[t]
\centering
\footnotesize
\begin{tabular}{lccc}
\hline
\textbf{Mission} & \textbf{\#test flights} & \textbf{JSD$_\text{speed}$} & \textbf{KL$_\text{spatial}$} \\
\hline
Spiral         & 2 & 0.32 & 7.71 \\
Grid           & 1 & 0.65 & 4.28 \\
Hover-transit  & 1 & 0.26 & 18.25 \\
Random (G--M)  & 1 & \textbf{0.16} & 18.87 \\
\hline
\textbf{Mean (across all test flights)} & 5 & 0.34 & 11.36 \\
\hline
\end{tabular}
\caption{Mobility-model fidelity on held-out AFAR flights, per mission. JSD on speed distributions and KL on spatial visit-frequency maps; lower is better.}
\label{tab:mobility_eval}
\end{table}

\textbf{The Gauss--Markov random walk produces the tightest velocity match (JSD~$=$~0.16) because its three free parameters $(\alpha, \bar v, \sigma_v^2)$ are fitted directly against the empirical speed distribution.} The deterministic mission models (spiral, grid, hover-transit) are fit against trajectory shape rather than the marginal speed distribution, so their JSD is higher (0.26--0.65)---this is expected: a fixed-speed boustrophedon scan cannot reproduce the entire empirical speed PDF, only its central tendency.

\textbf{Spatial KL is lower for area-coverage missions (grid: 4.28; spiral: 7.71) than for waypoint or random walks (hover-transit: 18.25; random: 18.87)} because area-coverage missions explore $(x,y)$ space uniformly by design, while waypoint and random walks concentrate occupancy mass around start/end points or local attractor regions, producing a more peaked spatial distribution that yields larger KL when compared to the empirical AFAR map.

\begin{figure*}[t]
\centering
\begin{subfigure}{\textwidth}
  \centering
  \includegraphics[width=0.95\textwidth]{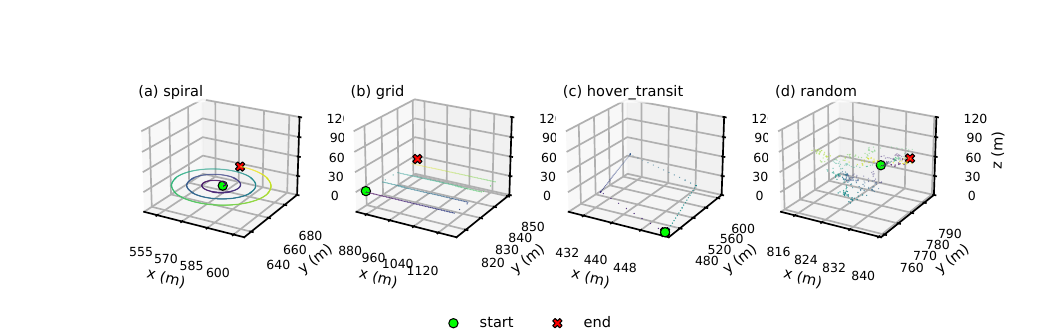}
  \caption{Simulated UAV trajectories per mission type.}
  \label{fig:mobility_traces}
\end{subfigure}

\vspace{1ex}

\begin{subfigure}{\textwidth}
  \centering
  \includegraphics[width=0.95\textwidth]{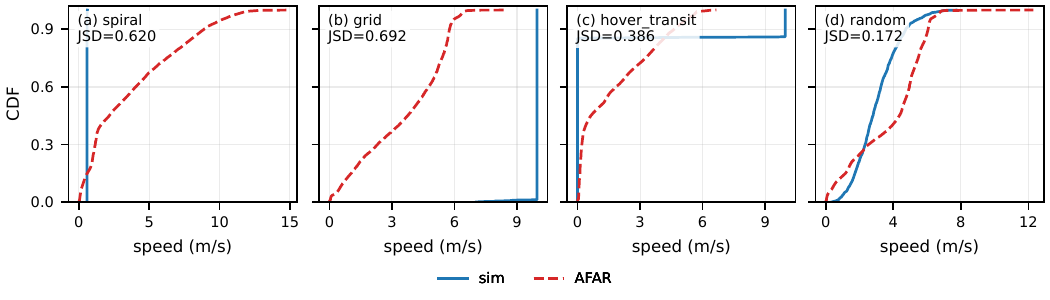}
  \caption{Velocity-distribution validation: simulated (dashed) vs.\ AFAR measured (solid) CDFs.}
  \label{fig:velocity_validation}
\end{subfigure}
\caption{Mobility-model calibration. (\subref{fig:mobility_traces})~Simulated UAV trajectories for the four mission types (spiral, grid, hover-transit, random). (\subref{fig:velocity_validation})~Per-mission velocity CDFs, simulated vs.\ AFAR measurements.}
\label{fig:mobility}
\end{figure*}

Figure~\ref{fig:mobility_traces} visualizes simulated 3D trajectories for each mission type. Figure~\ref{fig:velocity_validation} compares simulated vs.\ measured velocity CDFs.

\subsection{Link-Level and End-to-End Fidelity}
\label{sec:eval_e2e}

\textbf{The simulator reproduces its trajectory source's RSS distribution at Hellinger 0.33---well inside the 0.73 ceiling that separates two independent real-measurement campaigns.}

\textit{Three-way RSS fidelity.} We compare the RSS distributions of three sources on the log-distance$+$shadow variant: the Maeng measured campaign ($N=95$k, mean $-51.1$\,dBm), the AFAR digital-twin traces that drive the trajectories ($N=292$k, mean $-5.1$\,dBm), and our simulator ($N=1.8$M, mean $-24.5$\,dBm). The pairwise Hellinger Distances are $H(\mathrm{Maeng},\mathrm{Sim})=0.92$, $H(\mathrm{Maeng},\mathrm{AFAR})=0.73$, and $H(\mathrm{AFAR},\mathrm{Sim})=0.33$. The $0.73$ figure is the floor imposed by two real campaigns recorded with different hardware at different sites; our simulator matches its AFAR trajectory source at $0.33$, less than half that ceiling, while the larger Maeng--Sim distance reflects the campaigns' differing spatial sampling rather than a modeling error. Among the four calibration sweeps, log-distance$+$shadow attains the lowest total simulation-to-real divergence and is used throughout.

\textit{Link-level validation.} Replaying real AADM/AFAR trajectories through the digital twin, the calibrated log-distance variant attains a median absolute RSS error of 13.1\,dB (bias-corrected; 18.3\,dB raw, $-6.0$\,dB mean cross-sensor bias) over 1.8M paired samples, decreasing to 10.7\,dB at 60--80\,m altitude. The SNR time series correlates weakly but significantly with measurements (median Pearson $r \approx 0.09$, $p \approx 0$). Throughput RMSE is omitted: the source campaigns provide no measured end-to-end throughput.

\begin{figure}[t]
\centering
\includegraphics[width=0.95\columnwidth]{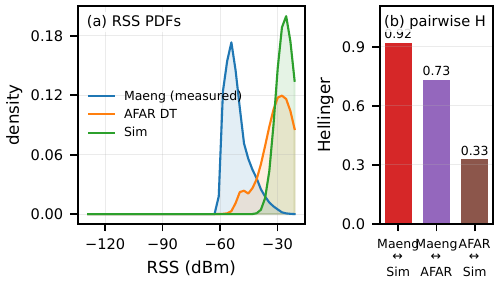}
\caption{Three-way RSS fidelity (Hellinger Distance, lower is better) for the log-distance$+$shadow variant.}
\label{fig:threeway}
\end{figure}

Figure~\ref{fig:threeway} presents the three-way Hellinger Distance comparison.

\begin{figure}[t]
\centering
\includegraphics[width=0.95\columnwidth]{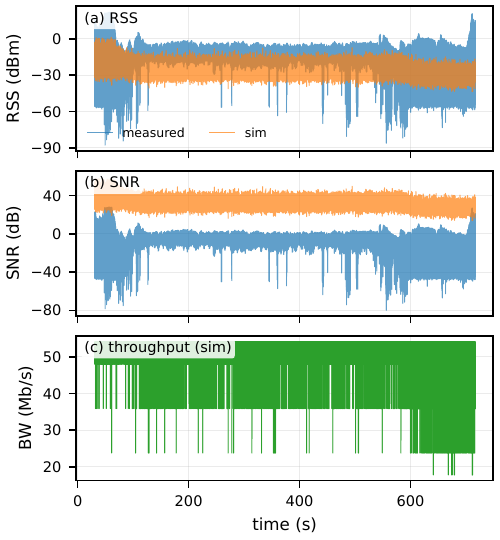}
\caption{Link-quality evolution for a representative flight (testbed~866, flight~23) under the log-distance$+$shadow variant: simulated vs.\ measured RSS and SNR over time, with simulated effective throughput.}
\label{fig:link_quality}
\end{figure}

Figure~\ref{fig:link_quality} shows SNR and effective bandwidth evolution over time for representative links during an AFAR trajectory replay.

\subsection{Dataset Diversity Analysis}
\label{sec:eval_diversity}

\textbf{\datasetname\ exhibits significantly higher statistical diversity than all existing benchmarks, confirming that the dataset captures genuine operational complexity rather than artificial noise.}

We quantify diversity using Hellinger Distance and Jensen--Shannon Divergence computed on inter-arrival time distributions, following the methodology of~\cite{netdiffusion2023}. Each flow is represented as a time series of inter-arrival packet times, as these are most affected by changing network conditions.

Table~\ref{tab:diversity} compares diversity across datasets.

\begin{table}[t]
\centering
\footnotesize
\begin{tabular}{llcc}
\hline
\textbf{Dataset} & \textbf{Attack} & \textbf{Hellinger} & \textbf{JSD} \\
\hline
\datasetname   & Blackhole+DDoS & 276.17 & 18.98 \\
               & DDoS+Replay    & 260.76 & 20.53 \\
               & DDoS           & 248.65 & 19.58 \\
               & DoS+Replay     & 167.83 & 15.30 \\
               & DoS            & 136.89 & 13.48 \\
               & DDoS+Wormhole  & 113.61 & 19.12 \\
               & DoS+Wormhole   & 81.49 & 15.39 \\
               & Blackhole+DoS  & 81.47 & 15.25 \\
               & Blackhole+Replay & 1.32 & 9.08 \\
               & Replay         & 1.32 & 9.09 \\
               & Replay+Wormhole & 1.30 & 8.91 \\
               & Wormhole       & 1.26 & 8.43 \\
               & Blackhole      & 1.20 & 7.96 \\
               & Blackhole+Wormhole & 1.19 & 7.88 \\
\hline
UNSW-NB15      & Backdoors      & 23.20 & 8.26 \\
               & Fuzzers        & 21.65 & 7.29 \\
               & Generic        & 21.53 & 6.46 \\
               & Reconnaissance & 21.46 & 7.23 \\
               & DoS            & 21.38 & 6.87 \\
               & Exploits       & 21.28 & 6.53 \\
               & Worms          & 21.14 & 7.22 \\
\hline
CICIOT2023     & DoS-UDP\_Flood & 21.44 & 13.93 \\
               & DDoS-UDP\_Flood & 11.07 & 9.29 \\
               & DDoS-UDP\_Fragmentation & 5.45 & 9.48 \\
               & DDoS-SlowLoris & 4.88 & 12.45 \\
               & DDoS-ICMP\_Flood & 4.59 & 9.47 \\
               & DoS-TCP\_Flood & 4.41 & 21.03 \\
               & DDoS-ICMP\_Fragmentation & 4.27 & 9.00 \\
               & DDoS-SYN\_Flood & 4.12 & 9.51 \\
               & Mirai-greip\_flood & 4.06 & 9.21 \\
               & Mirai-greeth\_flood & 3.97 & 9.14 \\
               & DDoS-RSTFINFlood & 3.89 & 8.52 \\
               & DoS-SYN\_Flood & 3.73 & 8.71 \\
\hline
UAV-NIDD       & DoS            & 27.71 & 17.53 \\
\hline
\end{tabular}
\caption{Statistical diversity comparison (Hellinger Distance / JSD) on inter-arrival time distributions.}
\label{tab:diversity}
\end{table}

\textit{The diversity structure of \datasetname\ is strongly bimodal and tracks the attack mechanism rather than mere attack presence.} High-rate attacks and their collaborative compositions separate sharply from benign traffic: Blackhole+DDoS reaches a Hellinger Distance of 276.17 and DDoS alone 248.65, far above the most diverse prior benchmark (UAV-NIDD DoS at 27.71, UNSW-NB15 Backdoors at 23.20, CICIOT2023 DoS-UDP\_Flood at 21.44). Stealth attacks, by contrast, sit close to benign: Blackhole (1.20), Wormhole (1.26), Replay (1.32), and their compositions all fall below 1.4 Hellinger, beneath every prior dataset. This is by design---blackhole, wormhole, and replay manipulate routing and packet timing while preserving benign-like inter-arrival statistics, so their low IAT divergence is the quantitative signature of their stealth, not a defect. The two-orders-of-magnitude dynamic range within a single dataset arises from genuine operational variability: the same attack executed at different UAV-to-base-station distances produces different flow signatures through wireless channel effects, and mobility-induced topology changes route the same attack along different paths.

\textbf{Internal diversity confirms that each configuration axis contributes measurably distinct flow distributions, ruling out the possibility that high cross-dataset divergence is an artifact of a single dominant axis.}

Table~\ref{tab:internal_diversity} reports the divergence between flow distributions induced by each pair of extreme axis values, holding all other axes mixed. A non-trivial value on every row demonstrates that no single axis dominates the dataset's diversity.

\begin{table}[t]
\centering
\footnotesize
\begin{tabular}{lc}
\hline
\textbf{Axis (low vs.\ high)} & \textbf{JSD} \\
\hline
5 vs.\ 20 drones                    & 13.18 \\
Log-distance vs.\ 3GPP path loss    & 8.30 \\
Spiral vs.\ random mission          & 8.58 \\
TX power 10 vs.\ 30\,dBm            & 8.65 \\
\hline
\end{tabular}
\caption{Internal diversity by configuration axis. Computed on \textbf{benign} flows to isolate the axis effect from class-mix confounding. Higher values indicate the axis produces measurably different flow distributions.}
\label{tab:internal_diversity}
\end{table}

\begin{figure*}[t]
\centering
\includegraphics[width=0.75\textwidth]{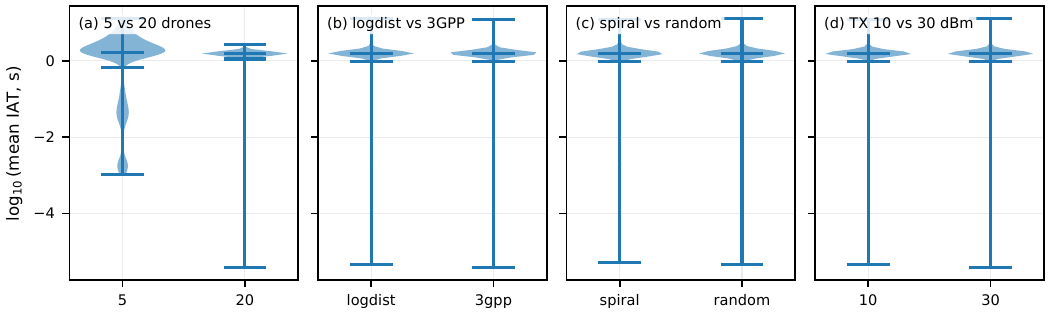}
\caption{Feature distributions across configuration axes: (a)~5 vs.\ 20 drones, (b)~log-distance vs.\ 3GPP, (c)~spiral vs.\ random, (d)~TX power 10 vs.\ 30\,dBm.}
\label{fig:internal_diversity}
\end{figure*}

\textit{Every configuration axis contributes non-zero divergence, confirming that the configurability of \datasetname\ produces genuine flow-level variability rather than nominal parameter changes.} On benign flows, swarm size dominates (5 vs.\ 20 drones, JSD 13.18), followed by transmit power (10 vs.\ 30\,dBm, 8.65), mission type (spiral vs.\ random, 8.58), and path-loss model (log-distance vs.\ 3GPP, 8.30). No single axis accounts for the dataset's diversity, which rules out the possibility that the high cross-dataset divergence in Table~\ref{tab:diversity} is an artifact of one over-weighted parameter. Figure~\ref{fig:internal_diversity} shows the per-axis feature distributions underlying these divergences.

\subsection{Attack Separability Analysis}
\label{sec:eval_separability}

\textbf{Attack families produce distinct but partially overlapping feature distributions, with stealth attacks (blackhole, wormhole) being hardest to separate from benign traffic.}

Table~\ref{tab:separability_stats} reports per-class summary statistics on three flow-level features that an IDS would commonly use. Each cell shows mean $\pm$ standard deviation across all flows of that class.

\begin{table}[t]
\centering
\footnotesize
\begin{tabular}{lccc}
\hline
\textbf{Class} & \textbf{IAT (ms)} & \textbf{Pkt rate (pps)} & \textbf{Pkt size (B)} \\
\hline
Benign     & 1,666 $\pm$ 642 & 0.63 $\pm$ 0.09 & 103.7 $\pm$ 26.6 \\
DoS        & 43.2 $\pm$ 50.9 & 176.0 $\pm$ 253.8 & 71.6 $\pm$ 13.6 \\
DDoS       & 219.5 $\pm$ 257.5 & 500 $\pm$ 8,570 & 74.2 $\pm$ 13.2 \\
Blackhole  & 1,647 $\pm$ 564 & 0.63 $\pm$ 0.17 & 104.4 $\pm$ 52.0 \\
Wormhole   & 1,671 $\pm$ 652 & 0.63 $\pm$ 0.08 & 103.4 $\pm$ 2.0 \\
Replay     & 1,890 $\pm$ 1,181 & 0.60 $\pm$ 0.12 & 102.6 $\pm$ 3.6 \\
\hline
\end{tabular}
\caption{Per-class summary statistics on key separability features. Mean $\pm$ standard deviation across all flows in \datasetname.}
\label{tab:separability_stats}
\end{table}

\begin{figure}[t]
\centering
\includegraphics[width=0.95\columnwidth]{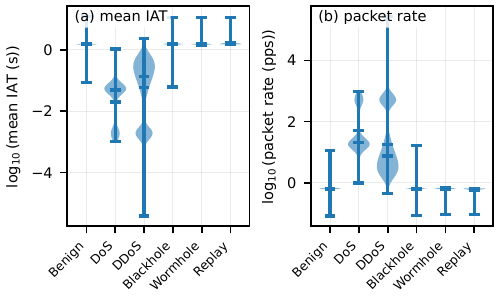}
\caption{Distribution of key flow features across attack families: (a)~mean inter-arrival time, (b)~packet rate, (c)~flow duration, (d)~mean packet size.}
\label{fig:attack_distributions}
\end{figure}

\textbf{Standard deviations on each feature are larger for stealth attacks (blackhole, wormhole) than for floods (DoS, DDoS), reflecting the design intent of stealth attacks to mimic the natural variability of benign UAV traffic.} This makes stealth attacks hard for static-feature classifiers and motivates the use of multi-feature and temporal-context detectors. Figure~\ref{fig:attack_distributions} shows the per-family distributions of these features.

\begin{figure}[t]
\centering
\includegraphics[width=0.85\columnwidth]{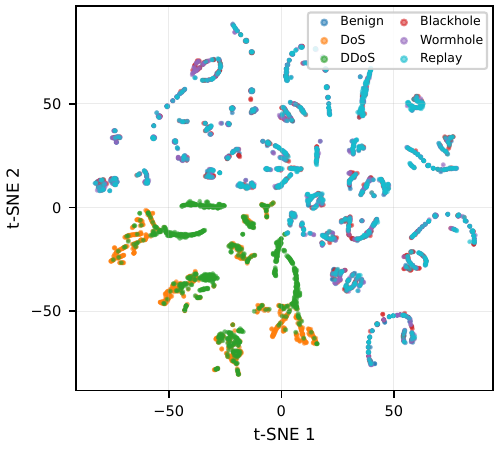}
\caption{t-SNE embedding of \datasetname\ flows colored by attack type. Stealth attacks (blackhole, wormhole) overlap with benign traffic.}
\label{fig:tsne}
\end{figure}

\textit{The embedding reflects the separability structure of Table~\ref{tab:separability_stats}.} Figure~\ref{fig:tsne} shows a t-SNE projection of the flow features. High-rate DoS and DDoS flows form compact clusters, set apart by their sub-second inter-arrival times and elevated packet rates, while benign flows spread broadly across the embedding. Blackhole, wormhole, and replay flows overlap the benign region, consistent with their near-benign inter-arrival, duration, and packet-size statistics. This overlap is the intended difficulty: stealth attacks are designed to blend with normal traffic, so a dataset in which they were trivially separable would be unrealistic. The embedding thus visualizes why \datasetname\ poses a genuinely hard detection problem rather than a separable toy task.

\subsection{Baseline IDS Benchmarking}
\label{sec:eval_baselines}

\textbf{All ten baselines saturate on binary attack detection across every dataset, including \datasetname, confirming the data is cleanly learnable; the genuine difficulty appears only in the full multi-class taxonomy.}

Table~\ref{tab:baseline_f1} reports binary Benign-vs-DoS weighted $F_1$ across four datasets (flow-statistics features, standard scaling).

\begin{table}[t]
\centering
\begin{tabular}{lcccc}
\hline
\textbf{Model} & \textbf{CICIDS} & \textbf{UNSW} & \textbf{CICIOT} & \textbf{\datasetname} \\
\hline
1D-CNN      & 97.31 & 98.71 & 93.14 & 99.98 \\
LSTM        & 92.15 & 87.95 & 83.38 & 98.86 \\
RF          & 99.92 & 99.11 & 97.57 & 99.99 \\
SGD         & 91.98 & 97.96 & 87.99 & 99.98 \\
LR          & 92.24 & 98.67 & 88.20 & 99.99 \\
MLP         & 97.35 & 98.65 & 92.41 & 99.98 \\
LightGBM    & 99.84 & 99.13 & 97.67 & 99.96 \\
ConvNet     & 96.59 & 98.48 & 91.30 & 99.99 \\
TinyML      & 99.89 & 98.86 & 97.07 & 99.98 \\
CNN-BiLSTM  & 95.63 & 98.70 & 94.35 & 99.99 \\
\hline
\end{tabular}
\vspace{1mm}
\caption{Binary Benign-vs-DoS weighted $F_1$ (\%) of IDS models across datasets (flow-statistics features, standard scaling). All architectures approach the learnability ceiling on \datasetname; min--max and robust scalers give closely matching results.}
\label{tab:baseline_f1}
\end{table}

\textbf{Binary detectability is a learnability floor, not the operational task; the difficulty is fine-grained attack identification.}

On binary Benign-vs-DoS detection every architecture saturates---nine of ten exceed $0.99$ weighted $F_1$ on \datasetname\ and the lowest (LSTM) reaches $0.989$---mirroring the high scores on the legacy benchmarks. This confirms the dataset is well-formed and learnable, but binary detection does not exercise the operational task. The discriminating signal lies in the full multi-class setting (Table~\ref{tab:per_attack}), where the same ten architectures must separate fifteen overlapping benign and attack classes. All three scalers (standard, min--max, robust) were evaluated and agree closely for most models, with the recurrent and hybrid architectures (LSTM, CNN-BiLSTM) the most scaler-sensitive.

\textbf{In the full 15-class native setting, per-class $F_1$ spreads enormously, and the low-rate stealth and replay classes are the primary source of difficulty.}

\begin{figure}[t]
\centering
\includegraphics[width=0.95\columnwidth]{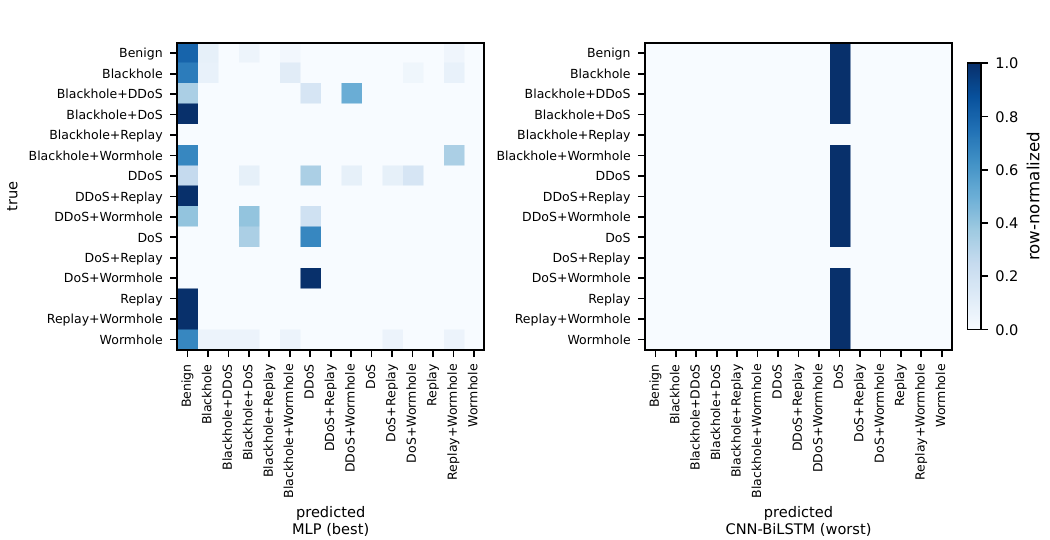}
\caption{Confusion matrix for the best-performing baseline on \datasetname.}
\label{fig:confusion}
\end{figure}

Table~\ref{tab:per_attack} reports per-class $F_1$ in the 15-class native multi-class setting on \datasetname\ (flow-statistics features, standard scaling); Figure~\ref{fig:confusion} shows the confusion matrices for the best- (LightGBM) and worst-performing (LSTM) baselines.

\begin{table*}[t]
\centering
\scriptsize
\setlength{\tabcolsep}{3pt}
\begin{tabular}{lccccccccccccccc}
\hline
\textbf{Model} & \textbf{Be} & \textbf{DoS} & \textbf{DDoS} & \textbf{Bh} & \textbf{Wh} & \textbf{Rp} & \textbf{Bh+Do} & \textbf{Bh+DD} & \textbf{Bh+Rp} & \textbf{Bh+Wh} & \textbf{DD+Rp} & \textbf{DD+Wh} & \textbf{Do+Rp} & \textbf{Do+Wh} & \textbf{Rp+Wh} \\
\hline
1D-CNN      & 22.03 & 15.53 & 40.35 & 7.11 & 7.79 & 3.24 & 5.67 & 16.24 & 0.72 & 4.32 & 10.34 & 11.81 & 5.03 & 7.09 & 0.79 \\
LSTM        & 12.80 & 5.85 & 6.30 & 8.59 & 0.00 & 0.81 & 0.91 & 0.00 & 0.16 & 1.57 & 3.64 & 9.52 & 0.94 & 2.93 & 0.07 \\
RF          & 42.91 & 36.17 & 47.77 & 21.45 & 24.42 & 7.55 & 37.09 & 58.15 & 3.05 & 20.87 & 28.58 & 61.05 & 28.64 & 39.42 & 6.86 \\
SGD         & 53.00 & 27.21 & 27.84 & 9.24 & 6.97 & 3.32 & 11.87 & 29.26 & 2.11 & 4.72 & 20.60 & 28.98 & 32.22 & 11.88 & 7.22 \\
LR          & 62.44 & 26.85 & 38.21 & 19.00 & 18.97 & 3.93 & 12.54 & 33.72 & 3.24 & 16.11 & 27.77 & 37.78 & 21.34 & 20.16 & 4.25 \\
MLP         & 18.42 & 10.56 & 17.79 & 12.64 & 5.61 & 3.16 & 2.54 & 11.03 & 1.00 & 4.95 & 11.04 & 8.18 & 2.91 & 4.70 & 0.99 \\
LightGBM    & 55.59 & 67.49 & 68.13 & 53.04 & 55.63 & 14.34 & 78.65 & 81.52 & 7.48 & 38.21 & 33.62 & 81.85 & 40.58 & 77.06 & 7.61 \\
ConvNet     & 31.46 & 11.56 & 45.65 & 16.19 & 15.33 & 4.05 & 2.31 & 12.37 & 0.77 & 4.42 & 11.61 & 13.29 & 3.66 & 8.18 & 1.65 \\
TinyML      & 51.17 & 36.66 & 52.91 & 32.24 & 36.32 & 9.05 & 37.98 & 62.79 & 3.55 & 28.70 & 27.57 & 59.50 & 16.95 & 35.38 & 4.26 \\
CNN-BiLSTM  & 40.67 & 4.15 & 9.48 & 4.54 & 0.00 & 2.53 & 0.51 & 2.72 & 0.27 & 0.57 & 2.61 & 20.17 & 1.63 & 1.40 & 0.42 \\
\hline
\end{tabular}
\vspace{1mm}
\caption{Per-class $F_1$ (\%) on \datasetname\ in the 15-class native multi-class setting.}
\label{tab:per_attack}
\end{table*}

\subsection{Collaborative Attack Detection}
\label{sec:eval_collab}

\textbf{Out-of-distribution generalization to unseen collaborative compositions is mixed: most compositions remain detectable, but blackhole-paired stealth combinations resist every baseline.}

To probe generalization, we train all ten baselines on single-attack flows only (compositions 1--7 in Table~\ref{tab:collab_taxonomy}) and test on the nine collaborative compositions (8--16). Table~\ref{tab:collab_f1} reports per-composition AUROC.

\begin{table*}[t]
\centering
\scriptsize
\setlength{\tabcolsep}{4pt}
\begin{tabular}{lcccccccccc}
\hline
\textbf{Composition} & \textbf{1D-CNN} & \textbf{LSTM} & \textbf{RF} & \textbf{SGD} & \textbf{LR} & \textbf{MLP} & \textbf{LightGBM} & \textbf{ConvNet} & \textbf{TinyML} & \textbf{CNN-BiLSTM} \\
\hline
Blackhole+DoS        & 96.73 & 77.32 & 96.42 & 92.96 & 95.67 & 94.26 & 98.60 & 98.50 & 92.61 & 95.66 \\
Blackhole+DDoS       & 97.01 & 80.93 & 93.40 & 72.24 & 94.26 & 96.92 & 99.69 & 98.58 & 92.59 & 93.61 \\
DoS+Wormhole         & 97.39 & 75.68 & 90.23 & 94.06 & 96.98 & 94.03 & 99.52 & 99.34 & 90.56 & 95.69 \\
DDoS+Wormhole        & 95.13 & 78.56 & 92.62 & 78.72 & 92.87 & 94.59 & 99.43 & 96.74 & 93.35 & 88.09 \\
Blackhole+Wormhole   & 60.67 & 52.65 & 85.80 & 50.75 & 51.31 & 70.95 & 84.97 & 57.44 & 86.37 & 69.14 \\
Blackhole+Replay     & 73.42 & 53.88 & 78.32 & 62.58 & 83.50 & 82.68 & 81.74 & 52.99 & 77.50 & 72.97 \\
DDoS+Replay          & 94.68 & 74.71 & 93.06 & 81.03 & 91.87 & 94.51 & 99.99 & 93.97 & 88.51 & 88.30 \\
DoS+Replay           & 94.36 & 75.08 & 94.69 & 87.19 & 93.29 & 92.22 & 99.58 & 94.78 & 89.48 & 94.74 \\
Replay+Wormhole      & 84.30 & 57.79 & 86.05 & 88.03 & 96.24 & 93.92 & 81.42 & 57.15 & 86.30 & 95.03 \\
\hline
\end{tabular}
\vspace{1mm}
\caption{Out-of-distribution AUROC (\%) on collaborative compositions. All three scalers agree closely; standard-scaler results are shown.}
\label{tab:collab_f1}
\end{table*}

\textit{Flood-bearing compositions are recovered well}---LightGBM, ConvNet, and TinyML exceed $0.93$ AUROC on every DoS- or DDoS-containing pairing, because the high-rate component dominates the flow signature even when the specific combination was never seen during training. The two blackhole$+$stealth compositions are the exception: Blackhole+Wormhole and Blackhole+Replay fall to $0.51$--$0.86$ across all ten models, since neither component perturbs flow statistics enough to be recognized out of distribution.

\textbf{The residual gap on stealth-paired compositions motivates dedicated multi-flow and context-aware detection.}

Single-flow classifiers process each flow independently and cannot exploit cross-flow correlations; when both components of a composition are individually near-benign, no single-flow feature suffices. Detecting these cases requires multi-flow features that capture network-wide behavior or temporal context across windows.

\subsection{Topology Characterization}
\label{sec:eval_topology}

\textbf{Mobility-induced topology changes produce measurable variation in flow statistics across simulation windows.}

\begin{figure}[t]
\centering
\includegraphics[width=0.98\columnwidth]{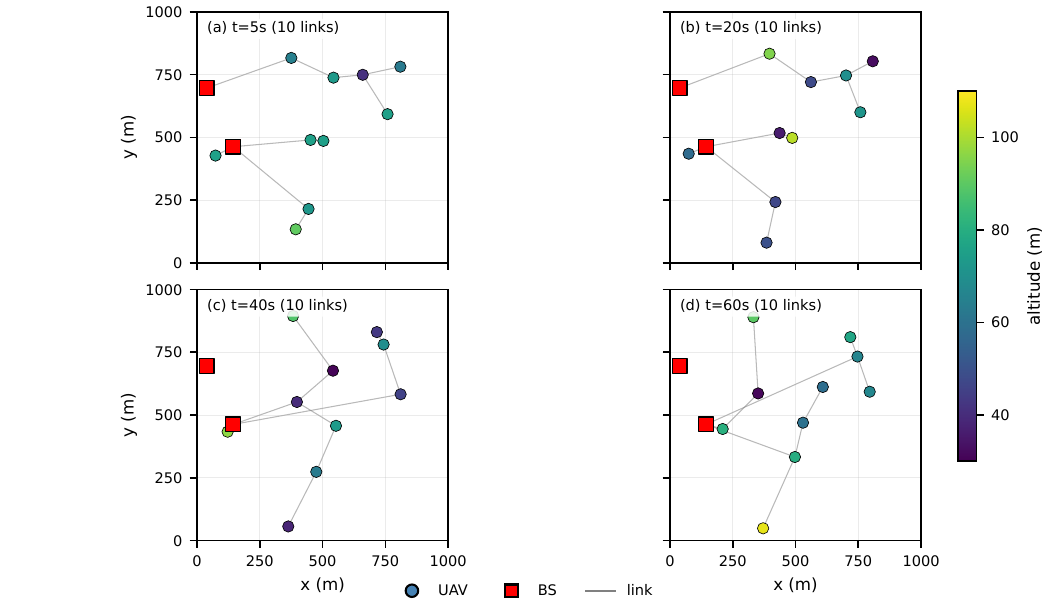}
\caption{Topology evolution for a 10-drone random-mission configuration (2 base stations, seed 42): snapshots at windows 1, 4, 8, and 12 ($t = 5, 20, 40, 60$\,s).}
\label{fig:topology}
\end{figure}

\textit{Figure~\ref{fig:topology} traces one configuration---a 10-drone random mission with two base stations---across four time windows.} Over 60\,s the swarm reorganizes substantially: individual UAVs traverse hundreds of meters, their altitudes cycle through the full 30--110\,m band (marker color), and each UAV's serving base station changes as it crosses the midline between \texttt{bs1} and \texttt{bs2}. The active-link count remains 10 throughout, but the \emph{identity} and \emph{length} of those links turn over from window to window, so the per-flow distance, path loss, and achievable rate that each flow experiences are continually re-sampled. This topology churn within a single fixed configuration is the mechanism behind the inter-window flow-statistic variation, and---aggregated over the 1{,}024 configurations---behind the dataset-level diversity quantified in Section~\ref{sec:eval_diversity}.

\section{Limitations and Future Work}
\label{sec:limitations}

We explicitly acknowledge the following scope limitations to guide appropriate use of \datasetname:

\textbf{Emulated wireless channel.} The digital twin applies wireless channel effects (delay, bandwidth, loss) via Linux \texttt{tc netem} inside Docker containers rather than through actual RF propagation. While our four-layer calibration pipeline validates fidelity against real AERPAW measurements, the emulation cannot capture fine-grained physical-layer phenomena such as fast fading, frequency-selective multipath, or Doppler shift. Path loss and correlated shadow fading are modeled; small-scale fading is not.

\textbf{TCP/IP protocol stack.} Benign traffic uses HTTP (nginx/curl) rather than MAVLink or other UAV-specific protocols. The traffic volume and burstiness patterns are realistic for sensing/telemetry payloads, but application-layer semantics differ from operational UAV protocols.

\textbf{No physical-layer attacks.} The five attack families target the network and routing layers. GPS spoofing, jamming, and other physical-layer attacks are not included.

\textbf{No energy model.} Battery drain and energy-constrained behavior are not simulated. UAV mission duration and resource allocation are fixed per configuration.

\textbf{Flow-level cap.} Flows are capped at 100{,}000 packets. For sustained high-rate attacks, this preserves rate-based features but truncates volume-based features. Downstream models should rely on rate and distributional features rather than raw packet counts.

\textbf{Single attack rate.} The campaign uses a single DoS/DDoS rate ($\sim$1{,}000\,pps). While this low-rate intensity is more operationally realistic than high-rate floods, future work should include multiple rate points to evaluate detection sensitivity across attack intensities.

\textbf{Future directions.} We plan to extend \datasetname\ with: (i)~fast fading and frequency-selective channel models, (ii)~MAVLink protocol integration, (iii)~GPS spoofing and jamming attacks, (iv)~larger swarm sizes (50--100 UAVs), (v)~heterogeneous UAV platforms with different radio capabilities, (vi)~multiple attack rate points, and (vii)~validation against additional AERPAW campaigns as new measurement data becomes available.

\section{Conclusion}
\label{sec:conclusion}

We presented \datasetname, a large-scale labeled network traffic dataset for intrusion detection in UAV swarm networks, generated from a Containernet-based digital twin systematically calibrated against NSF AERPAW testbed measurements. The digital twin employs a four-layer calibration pipeline---altitude-dependent path loss, mission mobility, link-level performance chain, and end-to-end trace fidelity---each validated against independent measurement campaigns. The resulting dataset spans 1{,}024 configurations across 5 attack families, 4 mission types, 2 path loss models, and 16 attack compositions including 9 complementary collaborative scenarios.

Our evaluation demonstrates four key findings. First, the calibration pipeline achieves 4.40\,dB overall RMSE path loss and 0.34 mean JSD mobility fidelity, and the digital twin reproduces its real trajectory source's RSS distribution at Hellinger 0.33---within the 0.73 cross-sensor ceiling that separates two independent real campaigns. Second, \datasetname\ exhibits significantly higher statistical diversity than all existing IDS benchmarks, with Hellinger scores up to an order of magnitude higher for high-rate attacks (276 vs.\ 28 for the most diverse prior dataset). Third, while ten standard baselines detect attacks near-perfectly in the binary setting (weighted $F_1$ above 0.98), their accuracy collapses in the full 15-class native taxonomy---per-class $F_1$ ranges from near zero to 0.82 and stealth attacks fall into the single digits---confirming that fine-grained attack identification under mobility-induced distributional shift is the real challenge. Fourth, when baselines trained only on single-attack flows are tested on unseen collaborative compositions, out-of-distribution generalization is mixed: most compositions remain detectable (AUROC above 0.9 for the strongest models), but blackhole-paired stealth combinations (Blackhole+Wormhole, Blackhole+Replay) stay difficult for all ten, indicating that these emergent patterns require dedicated multi-flow or temporal-context detection.

All code, calibration data, and the complete dataset are publicly released to enable reproducible research in UAV network security.

\bibliographystyle{ACM-Reference-Format}
\bibliography{references}
\end{document}